# Shaped-pulse optimisation of coherent soft-x-rays

R. Bartels*, S. Backus*, E. Zeek *, L. Misoguti*, G. Vdovin †, I.P. Christov ‡, M.M. Murnane*, H.C. Kapteyn*

*JILA, University of Colorado and National Institute of Standards and Technology, Campus Box 440, Boulder, CO 80309-0440*

*† Delft University of Technology, 2600 GA Delft, The Netherlands*

*‡ Department of Physics, Sofia University, Sofia, Bulgaria*

**High-harmonic generation is one of the most extreme nonlinear-optical processes observed to date.[1-3] By focusing an intense laser pulse into a gas, the light-atom interaction that occurs during the process of ionising the atoms results in the generation of harmonics of the driving laser frequency, that extend up to order ~300 (corresponding to photon energies from 4 to >500eV).[4,5] Because this technique is simple to implement and generates coherent, laser-like, soft-x-ray beams, it is currently being developed for applications in science and technology including probing of dynamics in chemical and materials systems[6] and for imaging.[7] In this work we demonstrate that by carefully controlling the shape[8] of intense light pulses of 6-8 optical cycles, we can control[9,10] the interaction of light with an atom as it is being ionised, in a way that improves the efficiency of x-ray generation by an order of magnitude. Furthermore, we demonstrate that it is possible to control the spectral characteristics of the emitted radiation and to "channel" the interaction between different-order nonlinear processes. The result is an increased utility of harmonic generation as a**

**light source, as well as the first demonstration of optical pulse-shaping techniques to control high-order nonlinear processes.**

This work builds on a number of other recent advances in laser and nonlinear-optical science and technology. The use of temporally-shaped ultrashort pulses to create "designer" atomic wavepackets[11], control two-photon absorption,[12,13] and to control molecular processes[11,14] has been a topic of increasing interest as laser technology has developed. Recent work has demonstrated techniques to increase the conversion efficiency of high-harmonic generation by using waveguide[15] and novel focusing geometries[16,17] and very short duration (≤25 fs, or ~10 optical cycles) light pulses.[18] Recent work has also shown that the temporal ordering of colours in the light pulse affects the spectral characteristics of the high-harmonic radiation.[19,20] However, that work was limited to simple adjustment of the light pulse "chirp," and did not reveal any ability to selectively control the high harmonic generation (HHG) process or to improve overall conversion efficiency. The present work uses a newly-developed laser system that generates high power ultrafast laser pulses with the capability of temporally reshaping them in a very precise and flexible manner.[21] We show that very subtle changes (resulting in changes in the time duration of the light pulse of only a few femtoseconds) can manipulate the electronic response of an atom, and thereby control the spectral characteristics and brightness of the soft x-ray coherent radiation. We have found this effect to be very general, occurring over a wide range of parameters and in different gases. In light of the past work, the effectiveness of this technique is unexpected. It illustrates significant differences between low-order nonlinear processes such as second-harmonic generation and high-order processes such as HHG that are inherently non-perturbative. Furthermore, under some sets of conditions it is possible to channel most of the high-harmonic emission into a single spectral peak creating a coherent, nearly-monochromatic soft x-ray source with a greatly enhanced brightness.

In the experiment, we focus light pulses from a kilohertz repetition-rate, millijoule-energy ultrafast laser system into a 175 μm diameter gas-filled capillary waveguide. The pulse peak intensity ($\approx 2 \times 10^{14}$ W/cm$^2$) is sufficient to field-ionise the atoms (typically argon), in the process generating high-harmonic XUV radiation. The capillary waveguide allows us to create an extended region of high laser intensity and long coherence length to efficiently generate the high-harmonic radiation,[15] which is observed using an x-ray CCD camera directly-coupled to a grazing incidence spectrometer.

The laser system incorporates a micromachined deformable mirror (MMDM) pulse-shaping apparatus.[22] This adaptive optic element is used in a position within the laser system where the frequency components of the light pulse are spatially separated and can change the relative path length of the various colours, thus reordering the arrival times of the colours of the short pulse. The light pulse bandwidth is ~80nm FWHM centered at 800nm, and is not altered by the pulse shaper. We recently demonstrated that this mirror, when combined with adaptive control using an evolutionary search procedure, can allow us to generate nearly perfectly transform-limited, 15 femtosecond (<6 optical cycles) light pulses from the laser.[21] In that case, we optimised the intensity of second-harmonic radiation (SHG) generated by the laser pulses passing through a thin frequency-doubling crystal.[23] The actual laser pulse shapes were characterized in both amplitude and phase using the established Frequency-Resolved Optical Gating (FROG) technique.[24] In the case of SHG-optimised control, the highest conversion efficiency was observed to correspond to the highest peak power obtainable for a given pulse spectrum (i.e. the time-bandwidth or fourier-transform limit of laser pulse, where all frequency components arrive at once). This is the "intuitive" result, since a nonlinear-optical process should be most efficient at the highest peak intensity.

For our experiments to optimise high-harmonic generation, we started with an 18fs, fourier transform-limited laser pulse obtained using the SHG optimisation technique. The initial curve of Figure 1 (iteration number 0) shows the HHG spectrum from argon gas around the 27th harmonic, using a transform-limited 18fs pulse. This curve also represents the highest overall signal level obtainable before learning optimisation of the HHG. We verified that adjustment of the linear chirp of the pulse did not improve the brightness of the HHG spectrum. To use the evolutionary strategy (ES) for HHG optimisation, a spectral window of 0.5nm surrounding the 27th harmonic was selected. The algorithm starts with a set of randomly-selected pulse shapes (corresponding to 19 random voltages on the pulse shaper), and searches for a pulse shape to maximize the signal intensity here while at the same time minimizing the adjacent 25th and $29^{th}$ order harmonics. The best (fittest) pulses are then randomly altered ("mutated") by adding a normally-distributed random variable to the value of each pixel, and retested. Each "parent" solution generates 5 mutated children. The algorithm converges to some optimal solution after approximately 50 iterations. Figure 1 shows the optimum solution after each iteration of the algorithm. Surprisingly, after optimisation the intensity of a single harmonic (27th) has been increased by a factor of 8, while adjacent harmonic orders are only slightly enhanced. The level of enhancement depends on the gas pressure, harmonic order, and gas type, with the largest enhancements (of over an order of magnitude) observed at a gas pressure corresponding to macroscopic phase-matching of the high-harmonic process, and for the harmonic ($27^{th}$) where neutral gas absorption of the light is near the minimum. However, significant (>3x) enhancement is observed under virtually every set of conditions. As different optimisation searches proceed, the relative heights of the various harmonics can change, and energy is channeled to the 27th harmonic as the search converges. The total integrated x-ray flux also increases by a factor of 2 – 6, depending on the gas pressure and harmonic orders being optimised. For example, in the case of Krypton gas, the combined nonlinearity and absorption characteristics of the gas result in ~6 harmonic peaks of comparable intensity before optimisation. Optimization of any particular peak in the range of 17-23 results in enhancement of this peak by a factor of 3-4x, while harmonics immediately adjacent to the "target" peak see enhancement of ~1.5-2x, and other peaks see negligible or negative enhancement.

Figure 2 shows the laser pulse shapes corresponding to the transform-limited and final HHG spectra shown in Fig. 1. The transform-limited and optimised laser pulse shape are shown in Fig. 2a, while Fig. 2b shows the corresponding phase values. The optimised pulse has more structure and phase change on the leading edge, but is not much broader than the transform-limit (~21.6fs compared with 18fs). Thus, very slight modification of the driving pulse can result in drastic changes in the harmonic radiation generated. We have found that different optimization runs, even using the same fitness criteria, yield slightly differing pulse shapes even when similar enhancements are observed, suggesting that some degrees of freedom for the pulse shape are not significant, while others are; work is in progress to modify the fitness criteria to try to identify the critical features. The enhanced prepulse observed in the optimised pulse is not always present, and is likely a result of limitations of the phase-only pulse shaper; however, the presence of some phase structure on the leading edge of the pulse is a consistent feature. Furthermore, since the intensity of the prepulse is below that required for ionization and therefore HHG emission, it is unlikely to be significant. By selecting different criteria (i.e. different spectral windows, different harmonics, and/or different gases), the ES search algorithm can converge on different optimal solutions. Figures 3, for example, shows the results before (blue) and after (red) optimisation of the 27th harmonic in argon, but in this case optimised over a shifted spectral window positioned to the left of the harmonic peak, with no suppression of adjacent peaks. The optimisation process increased both the peak intensity and spectral purity of the harmonic, increasing the peak by over an order of magnitude. Choice of the optimization "window" on the red or the blue side of the initial peak consistently results in a red or blue shift of the harmonic; however, the resulting spectral bandwidth varies. This optimization is consistent with previous work that showed a positive (red to blue) chirp on the driving pulse can result in both a red shift and a spectral narrowing of the harmonic emission peaks[19] (although in that case no dramatic increase in brightness or selectivity was observed). The measured width of the optimised 27th harmonic is 0.24eV, corresponding to our instrument resolution. Thus, the true peak and spectral enhancements may be even higher. Since the duration of the XUV pulse is likely somewhat shorter than the driving laser pulse, the resulting HHG spectrum corresponds to an XUV pulse much closer to the time-bandwidth limit than before optimisation. This result illustrates quite clearly that pulse shaping can alleviate what had been thought to be a fundamental "tradeoff" for use of HHG as a light source—that the use of very short driving pulses, although dramatically increasing the efficiency of HHG, results in a broader spectrum for the individual harmonic peaks.

To explain the unexpected result that very slight changes in laser pulse shape can dramatically enhance and select individual harmonics, we consider the semi-classical rescattering model of HHG.[25,26] From a classical point of view, electrons are ionised and accelerated away from the core during one half-cycle of the laser field, and can be subsequently driven back to the core when the laser field reverses. Some fraction of the ionised electrons can recombine with the parent ion and give off their energy in the form of high harmonics. This process occurs over a few optical cycles of the laser field, resulting in a ≈ 5fs x-ray burst. Particular harmonics are generated by electrons returning to the core with a specific return energy that is related to the exact time within the optical cycles when the electron is initially ionised. We believe that the adaptive optimisation algorithm finds a pulse with the correct phase sequence to ensure that the continuum x-ray emission generated by a particular cycle of the laser pulse reinforces constructively or destructively with different parts of the continuum generated by adjacent cycles. This coherent control can lead to channeling and redirection of energy between different high-order nonlinear interactions. These findings therefore demonstrate a new type of intra-atomic "phase matching" between the laser field and the wavefunction of the ionised electron. Preliminary calculations based on phase-only pulse shaping confirm this interpretation. From a quantum point of view, our optimised laser pulse can adjust the quantum phase of the electron wavefunction which returns to the core, to optimise it for a particular harmonic feature.

In summary, this represents the first demonstration of coherent control of very a high-order nonlinear processes with a result that has immediate utility. The technique is considerably more effective than expected and illustrates very significant differences between low-order nonlinear processes such as second-harmonic generation and (inherently nonperturbative) high-order processes. We have also demonstrated selective channeling of energy from one high-order nonlinear process to another, revealing a new type of intra-atomic phase-matching. This work may lead to other new methods for control of highly nonlinear systems, as well as improving the utility of the HHG source for application experiments in a broad range of science.[27-30]

**Acknowledgements**

The authors gratefully acknowledge support for this work from the Department of Energy and the National Science Foundation.

Correspondence and requests for materials should be addressed to H.C.K. (e-mail: kapteyn@jila.colorado.edu)

**Figure Captions**

**Figure 1:** Optimisation of a single (27$^{th}$) harmonic in argon while suppressing adjacent harmonics. The optimisation criterion (fitness function) for this run corresponded to f = $\Sigma s_j$-$\Sigma s_i$-$\Sigma s_k$, where s corresponds to the signal level of a CCD pixel, and the ranges i, j, and k correspond to the pixels corresponding to a 0.5nm spectral bandwidth centered around the 25$^{th}$,27$^{th}$, and 29$^{th}$ harmonic. The peak enhancement for the 27$^{th}$ harmonic is a factor of 8, while the energy enhancement is a factor of 4.6. The contrast ratio between the 27$^{th}$ harmonic and adjacent harmonic increases by a factor of 4.

**Figure 2:** Amplitude (a) and phase (b) of the transform-limited (dashed) and optimised (line) laser pulses corresponding to Fig. 1. These data were taken using the frequency-resolved optical gating technique (FROG). The energy of the pulse is kept constant.

**Figure 3:** Optimisation of a single harmonic in argon with a spectral window at longer wavelengths than in Fig. 1 and without suppressing adjacent harmonics i.e. the fitness function is f = $\Sigma s_j$, where the range j corresponds to pixels to the "red" side of the 27$^{th}$ harmonic. The harmonic peak is enhanced by over an order of magnitude. Harmonics before and after optimisation are shown in blue and red respectively.

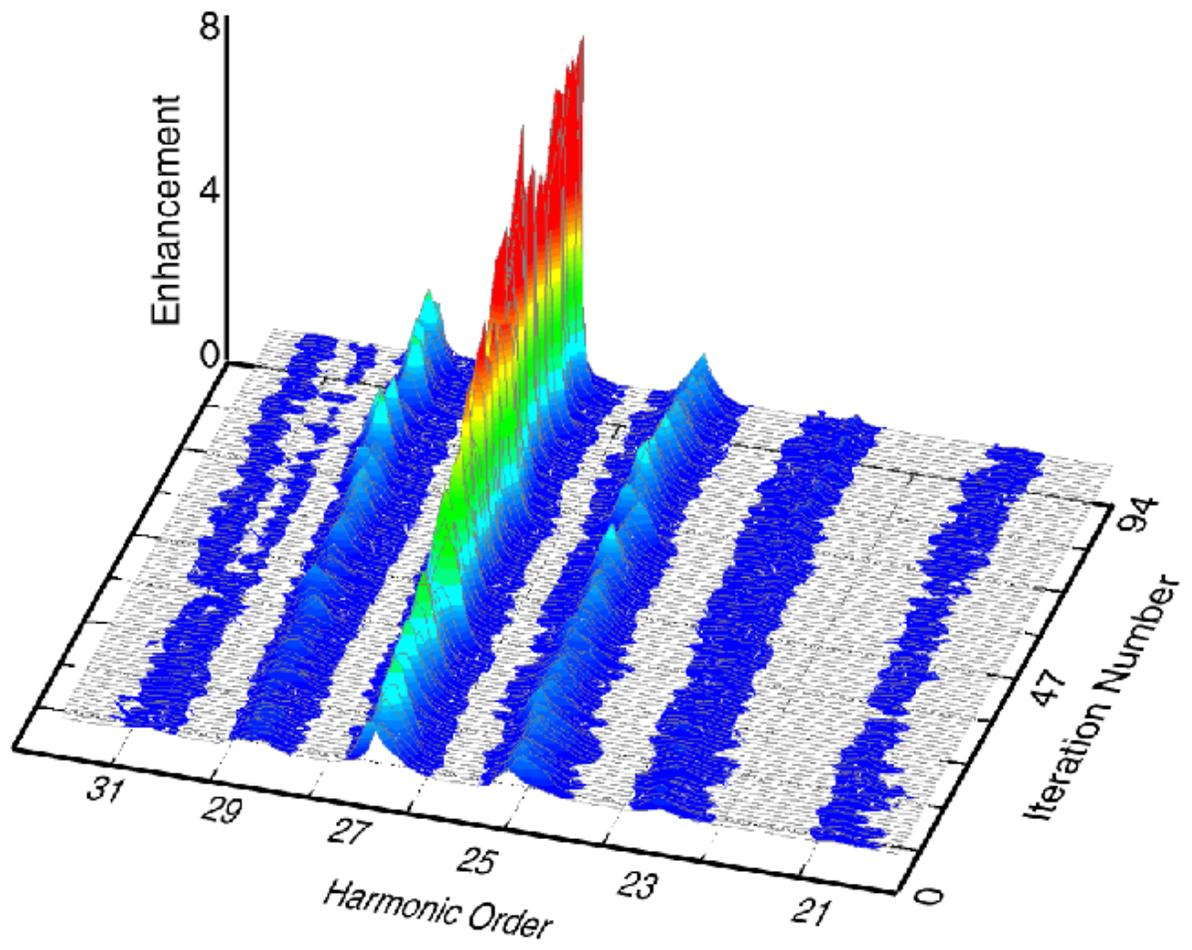

Figure 1; Bartels, R. et. al.

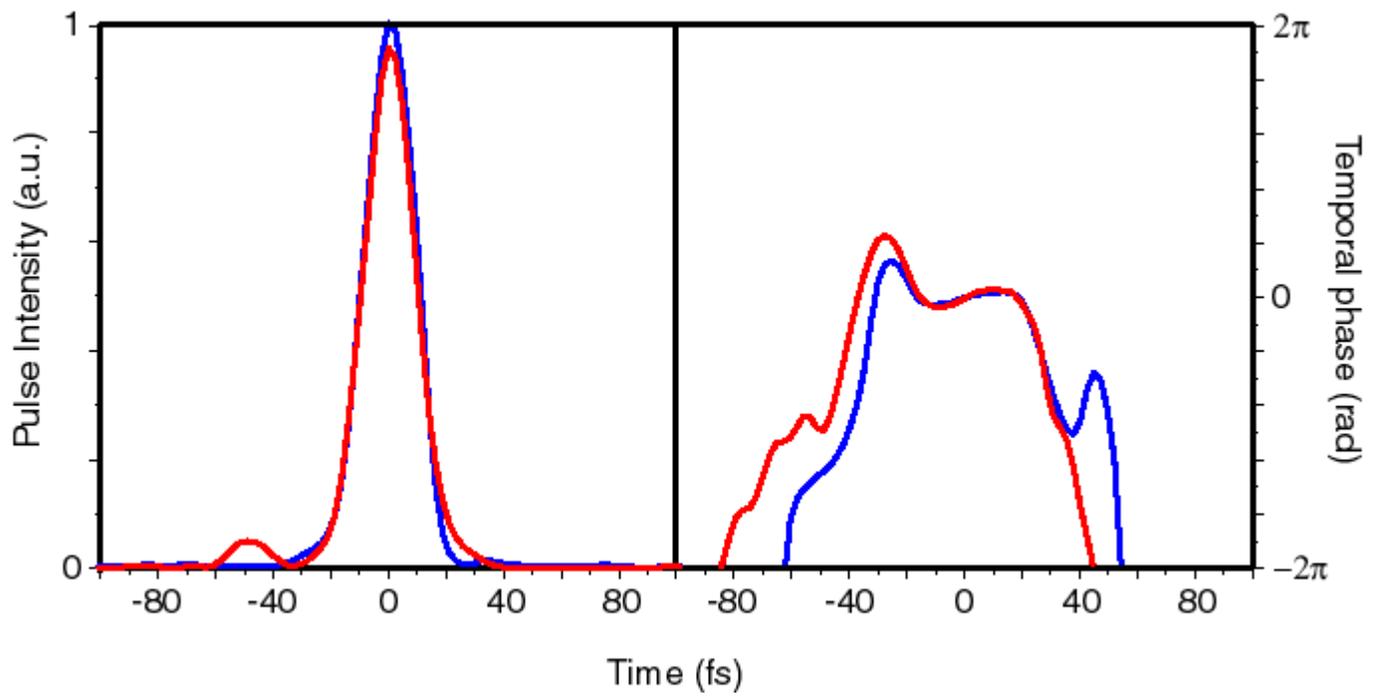

Figure 2; Bartels, R. et. al.

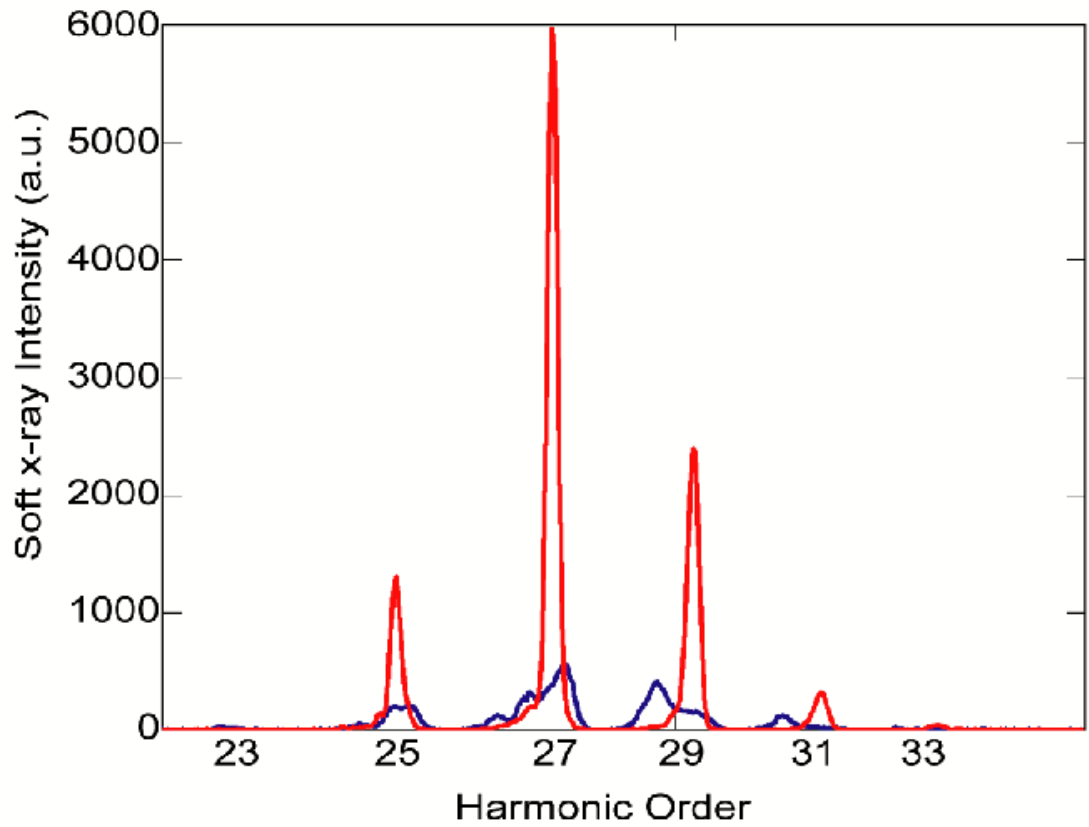

Figure 3; Bartels, R. et. al.